\documentclass{emulateapj}

\usepackage{natbib}
\usepackage{psfig}

\shorttitle{AGN SEDs}
\shortauthors{A. Alonso-Herrero et al.}

\begin{document}


\title{The nature of luminous X-ray sources 
with mid-infrared counterparts}



\author{A. Alonso-Herrero\altaffilmark{1,2}, 
P. G. P\'erez-Gonz\'alez\altaffilmark{1}, 
J. Rigby\altaffilmark{1}, 
G. H. Rieke\altaffilmark{1},
E. Le Floc'h\altaffilmark{1}, 
P. Barmby\altaffilmark{3}, 
M. J. Page\altaffilmark{4}, 
C. Papovich\altaffilmark{1}, 
H. Dole\altaffilmark{5,1},  
E. Egami\altaffilmark{1}, 
J.-S. Huang\altaffilmark{3},
D. Rigopoulou\altaffilmark{6},
D. Crist\'obal-Hornillos\altaffilmark{7},
C. Eliche-Moral\altaffilmark{7},
M. Balcells\altaffilmark{7},
M. Prieto\altaffilmark{7},
P. Erwin\altaffilmark{7},
C. W. Engelbracht\altaffilmark{1}, 
K. D. Gordon\altaffilmark{1},
M. Werner\altaffilmark{8},  
S. P. Willner\altaffilmark{3}, 
G. G. Fazio\altaffilmark{3}, 
D. Frayer\altaffilmark{9}, 
D. Hines\altaffilmark{1}, 
D. Kelly\altaffilmark{1},
W. Latter\altaffilmark{9}, 
K. Misselt\altaffilmark{1}, 
S. Miyazaki\altaffilmark{10},
J. Morrison\altaffilmark{1}, 
M. J. Rieke\altaffilmark{1},
G. Wilson\altaffilmark{9}}

\altaffiltext{1}{Steward Observatory, The University of Arizona, 
Tucson, AZ 85721; e-mail: aalonso@as.arizona.edu}
\altaffiltext{2}{IEM, 
CSIC, E-28006 Madrid, Spain}
\altaffiltext{3}{Harvard-Smithsonian Center for Astrophysics, 
Cambridge, MA 02138}
\altaffiltext{4}{Mullard Space Science Laboratory, University 
College London, Dorking, Surrey, RH5 6NT, U. K.}
\altaffiltext{5}{Institut d'Astrophysique Spatiale, b\^at 121, Universit\'e
Paris Sud, F-91405 Orsay Cedex, France}
\altaffiltext{6}{Department of Astrophysics, Oxford University, 
Keble Rd, Oxford, OX1 3RH, U. K.}
\altaffiltext{7}{Instituto de Astrof\'{\i}sica de 
Canarias, E-38200 La Laguna, Spain}
\altaffiltext{8}{Jet Propulsion Laboratory, 
Caltech, Pasadena, CA 91109}
\altaffiltext{9}{Spitzer Science Center, Caltech, Pasadena, CA 91125}
\altaffiltext{10}{Subaru Telescope, National Astronomical 
Observatory of Japan, Hilo, HI 96720}



\begin{abstract}
We investigate the luminous X-ray sources in
the Lockman Hole (LH) and the Extended Groth Strip (EGS) 
detected at  $24\,\mu$m 
using MIPS and also with IRAC on board {\it Spitzer}.
We assemble optical/infrared spectral energy distributions (SEDs) for 
45 X-ray/$24\,\mu$m 
sources in the EGS and LH. Only about 1/4 of the hard X-ray/24$\mu$m 
sources show pure type 1 AGN 
SEDs. More than half of the X-ray/$24\,\mu$m sources have 
stellar-emission-dominated or obscured SEDs,  similar to those of local 
type 2 AGN and spiral/starburst galaxies. 
One-third of the sources detected in hard X-rays do not
have a $24\,\mu$m counterpart. Two such sources in the LH have SEDs 
resembling those of S0/elliptical galaxies. 
The broad variety of SEDs in the optical-to-{\it Spitzer}
bands of X-ray selected AGN means that AGN selected according to the behavior in 
the optical/infrared will have to be
supplemented by other kinds of data (e.g., X-ray) to produce unbiased samples of AGN.  
\end{abstract}




%
\section{Introduction}
The very rapid evolution of quasars from $z \sim 2$  to the present
(Boyle et al. 1987) 
raises the question of whether any properties of these sources other
than space density have changed over this interval. One way to probe 
changes in the AGN population is to compare spectral energy distributions (SEDs)
over a broad frequency range as a function of redshift. The
SED of an active galaxy reflects the presence of the underlying AGN, plus the luminosity
of the host galaxy stellar population, the reddening of the AGN, and
the role of star formation, all in different frequency regimes. SED determination
in large samples of high-$z$ AGN using imaging detectors is
therefore an efficient way to survey for evolutionary trends in quasars
and AGN, and in their host galaxies.

A number of trends might be expected. Will we
be able to confirm predictions that the relative number of obscured AGN was higher
at large lookback times than now, to fit models of the X-ray background
(e.g., Gilli, Risaliti, 
\& Salvati 1999)? AGN activity may be
triggered by gas inflow caused by galaxy interactions.
Interactions also trigger starbursts --- will we find that an elevated
rate of star formation in the host galaxy is a typical characteristic
of distant AGN? Will younger AGN host galaxies have more 
material in their ISM than
current-epoch ones, causing greater extinction of their nuclei?

To probe such questions, we have used 
IRAC (Fazio et al. 2004), MIPS
(Rieke et al. 2004), and ancillary data to assemble SEDs for
 AGN in the Lockman Hole (LH) and
Extended Groth Strip (EGS). We have made use of deep 
X-ray images to locate the AGN. For this initial survey, 
we wanted an unambiguous
detection of an infrared (IR) excess to make SED 
classification robust.  We have included objects detected at
$24\,\mu$m, although we briefly discuss galaxies not detected at $24\,\mu$m.  
We compare the results with SEDs of a sample of nearby ($z < 0.12$) 
hard X-ray selected AGN that
are bright in the mid-IR, and hence nominally similar to the
sources identified in the deep {\it Spitzer}/X-ray fields. 
This comparison allows us to make tentative identifications of trends
in the AGN/host galaxy behavior from the typical redshift of the AGN
in the survey fields ($z \sim 0.2$ to 1.6) to the present.


\section{Spitzer  Observations} 

We have obtained  $24\,\mu$m MIPS observations 
of two fields in the LH: primary field (area of $5\arcmin \times 5\arcmin$)
at ${\rm RA}=10^{\rm h}\, 
52^{\rm m}$ and ${\rm Dec}= 57 \degr \,25\arcmin$ (J2000),
and parallel field (area of $7\arcmin \times
6\arcmin$)  at 
${\rm RA} = 10^{\rm h} \,52^{\rm m}$ and ${\rm Dec}=57\degr \,37\arcmin$. 
We also obtained
MIPS observations of the EGS 
overlapping with the {\it Chandra} observation ($\simeq 180\,$arcmin$^2$,
see next section)
at ${\rm RA}=14^{\rm h}\, 
17^{\rm m}$ and ${\rm Dec}= 52 \degr \,28\arcmin$. 
Gordon et al. (2004), Egami et al. (2004), Papovich et al. (2004),  
and Le Floc'h et al. (2004) describe 
the data reduction and photometry in detail. 
The astrometric uncertainties of {\it Spitzer} observations 
are $<1\arcsec$.
The 80\% completeness limits at $24\,\mu$m are: $0.17\,$mJy
and $0.1\,$mJy for the LH primary and parallel fields,
and $0.11\,$mJy for the EGS (Papovich et al. 2004).

The LH primary field and the EGS were  observed by IRAC 
at 3.6, 4.5, 5.8, and $8\,\micron$.
The data reduction is discussed in Huang et al. (2004) for the LH 
and Barmby et al. (2004, in preparation) for the EGS. 
Counterparts of all MIPS and X-ray sources were nearly pointlike, 
and the photometry used circular apertures and the standard
point-source calibration (see Huang et al. 2004).

\section{X-ray Observations}

The {\em XMM} images of the LH  
were formed from seven datasets 
taken in 2000 and 2001 to a total integration time of  
150\,ks.  The data cover entirely both the primary and parallel
$24\,\mu$m MIPS fields. 
We produced images in the energy
bands $0.2-0.5\,$keV, $0.5-2\,$keV, $2-5\,$keV, and $5-10\,$keV,  
and searched for sources simultaneously
in an iterative process to  optimize the
background model and thereby the sensitivity. 
In the LH we have detected 35 XMM sources with 
X-ray fluxes down to $f_{\rm 0.2-10keV} \simeq 10^{-15}\,{\rm erg\, cm
}^{-2\,}{\rm s}^{-1}$. The astrometric 
uncertainties are better than 1\arcsec \ 
for bright sources, and less than $3\arcsec$ \  for the faintest sources.

For the EGS, three {\it Chandra} ACIS datasets were taken from the CXC archive
with an exposure time of 131\,ks.  We
searched for sources separately in four bands ($0.5-8\,$keV, $0.5-2\,$keV, 
$2-8\,$keV, and $4-8\,$keV) using {\sc wavdetect}. 
In the overlapping area  between the {\it Chandra} 
and  MIPS fields
we have detected  77 sources in the full band 
with fluxes down to $f_{0.5-8{\rm keV}} \simeq 10^{-15}\,{\rm erg\, cm
}^{-2\,}{\rm s}^{-1}$. Of these, 
40 are detected in the $2-8\,$keV band. Astrometric 
uncertainties are $1-2$\arcsec, where the high value is for sources at large 
off-axis angles.

\section{Cross-correlation of 
X-ray and $24\,\mu$m sources}
Taking into 
account the astrometric uncertainties we used radii
of 2.2\arcsec \ and 3\arcsec \ for matching $24\,\mu$m sources 
to Chandra and XMM sources, respectively.
Within the LH approximately $57$\%  of the XMM sources 
are detected at $24\,\mu$m. 
75\% of LH X-ray sources with $f_{\rm 5-10keV} > 
10^{-15}\,{\rm erg\, cm}^{-2}\,
{\rm s}^{-1}$ have $24\,\micron$
counterparts. In the EGS
approximately 50\% of  Chandra sources have a 
$24\,\mu$m counterpart, and this fraction is $\simeq 60\%$ for  
sources detected in the hard ($2-8\,$keV) band.
Taking into account the surface density of
$24\,\mu$m sources (80\% completeness limit), the probability of a 
chance match with an X-ray source is $2-3$\%  for both the LH
and EGS.

\begin{figure}
\plotone{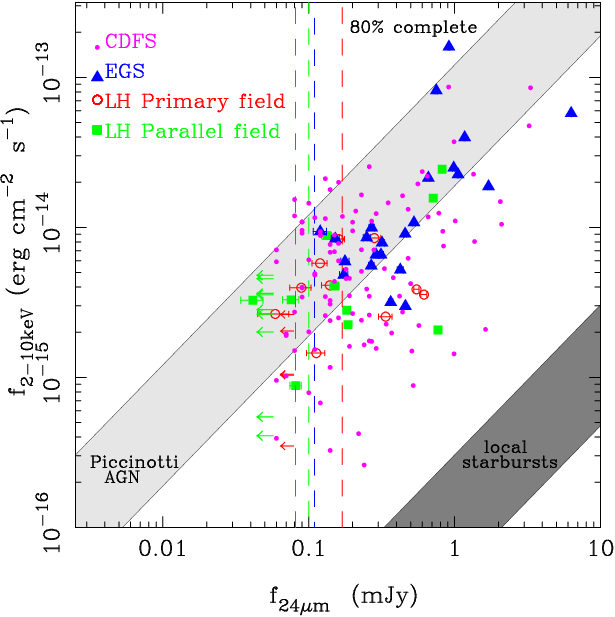}
\caption{$24\,\mu$m fluxes vs. $2-10\,$keV 
X-ray fluxes: XMM sources 
(squares and open circles) in the LH and 
Chandra sources in the EGS (filled triangles). 
The Chandra $2-8\,$keV fluxes have been converted 
to $2-10\,$keV assuming a power law with photon
index $\Gamma = 1.4$.
Also shown are X-ray/$24\,\mu$m
sources in the CDF-S (dots, Rigby et al. 2004). The $24\,\mu$m
non-detections of LH X-ray sources are shown as upper limits at 
a $5\sigma$ confusion limit.
The dashed lines are the 80\% completeness limits for the 
different fields (color coded as the symbols for
the different fields).
The lightly shaded area is the extrapolation of the median hard X-ray to mid-IR 
ratios ($\pm 1\sigma$) of local ($z<0.12$) hard X-ray selected
AGN (Piccinotti et al. 1982) with mid-IR emission. 
The dark shaded area is the extrapolation of local starburst galaxies from 
Ranalli, Comastri, \& Setti (2003).}
\end{figure} 


\begin{figure}

\plotone{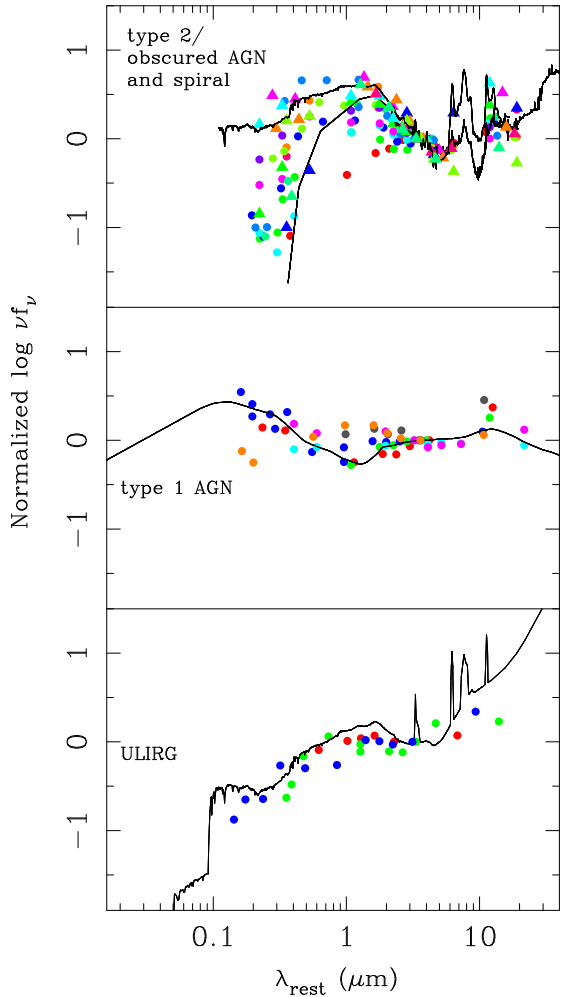}
\caption{SED (normalized at $
\lambda_{\rm rest} \simeq 3.5\,\mu$m) 
types for EGS X-ray/$24\,\micron$ sources 
(filled circles or triangles
of different colors, for clarity; Table~1)
with photometric redshifts estimated by us 
or redshifts from the literature.
The ULIRGs SED class is shown 
with the Mkn~273 template from Devriendt et al. (1999);
the type 1 AGN SED class is shown with the median QSO 
of Elvis et al. (1994). 
Obscured AGN/type 2 AGN/spiral SED galaxies are intermediate
between the Circinus template (lower one) and the 
M82-like template (upper one) of Le Floc'h et al. (2004).} 
\end{figure}

\section{Activity classification}

Hard X-ray   to mid-IR flux ratios are known to be different 
for AGN dominated galaxies and starbursts
in the Local Universe,  and 
can be used to assess
if the AGN emission is dominant in the mid-IR.
Fig.~1 shows the $24\,\mu$m fluxes vs.
$2-10\,$keV X-ray fluxes for our sample. 
As a comparison  we plot the extrapolation to 
fainter fluxes of the region occupied by hard X-ray selected AGN 
from Piccinotti et al. (1982) with detected mid-IR emission and 
$z<0.12$. This sample should be nominally similar to the
sources studied here.
The effect of increasing 
redshift on the observed ratio of hard X-ray to mid-IR fluxes is small for AGN
with low X-ray column densities, but it will make this ratio increase for 
Compton thick AGN at higher $z$ (Alexander et al. 2001). For starbursts
at $z<1$, Alexander et al. (2001) predict just a slight decrease of 
the ratio of hard X-ray to mid-IR emission for increasing $z$. The 
majority of sources in this study appear to derive their X-ray emission
from powerful AGN because they lie in the region of Fig.~1 populated by local
hard X-ray-selected AGN  extrapolated to fainter 
fluxes (Fadda et al. 2002; Franceschini et al. 2001; 
Alexander et al. 2001), and because for our sample  the $z$-dependent effects 
(see next section) in this figure are small.  

Also shown in Fig.~1 are 
sources in the Chandra Deep Field South (CDF-S) from Rigby et al. (2004). The 
greater sensitivity of the CDF-S X-ray data (compared with the EGS and LH) results
in detection of relatively weaker X-ray sources that either have a greater portion
of their luminosity generated from star formation, or are obscured in the X-rays
(compare to figure~1 in Alexander et al. 2002). Indeed,
spectroscopic observations of faint sources detected in the deepest X-ray surveys to date
indicate that these are starbursts  and low-redshift normal galaxies
(Barger et al. 2003).

\begin{figure}

\plotone{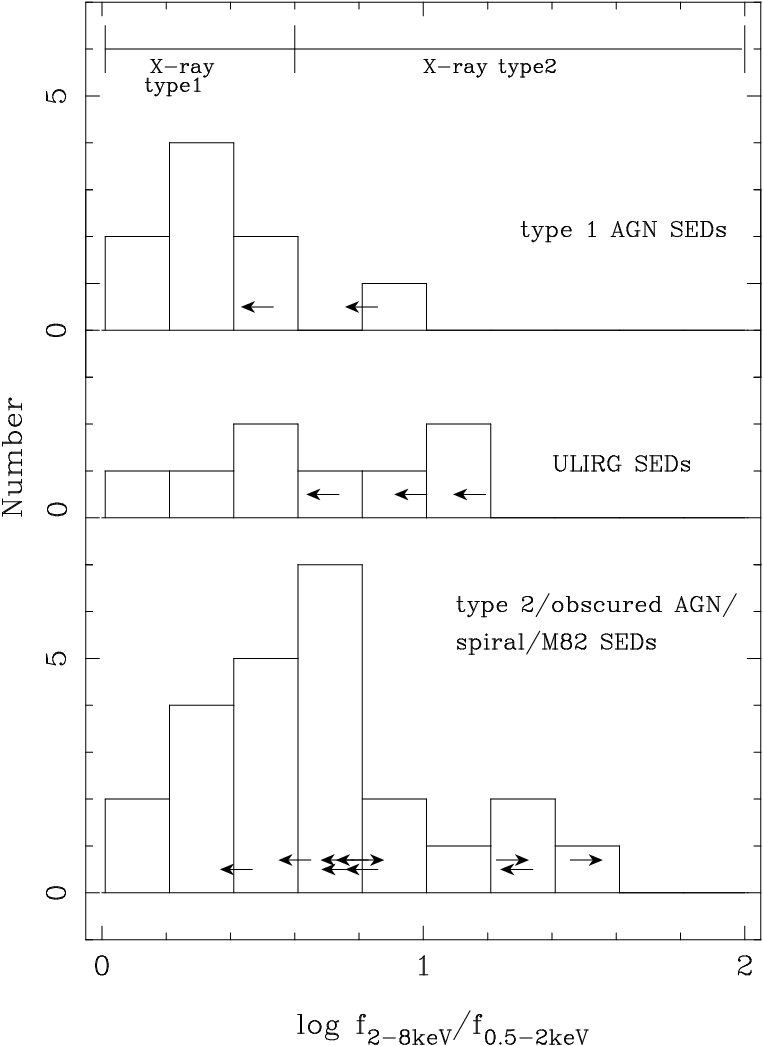}
\caption{Hard to soft flux ratio distributions 
for the SED types (Fig.~2) of
bright X-ray/$24\,\mu$m sources in the LH and EGS. 
The approximate division between type 1 and type 2 AGN
is also shown (Rigby et al. 2004 based on Szokoly et al. 2004). 
The arrows inside the histogram bins indicate sources with upper or
lower limits to the hard to soft flux ratios.}
\end{figure} 

All X-ray sources 
in the LH not detected at $24\,\mu$m  appear to be consistent
with being type 1 AGN or S0/elliptical galaxies (see 
next section). Alexander et al. (2002) found that most X-ray emitting 
galaxies in the HDF-N with no mid-IR emission 
are classified spectroscopically as S0/elliptical galaxies; that is,
they showed no emission-line evidence for activity.

X-ray hardness ratios can be used to distinguish between 
low-obscuration (soft)  and
high-obscuration (hard) AGN (e.g., Hasinger et al. 2001; Mainieri et al. 
2002; Szokoly et al. 2004). This X-ray classification 
agrees with the spectroscopic classification of
type 1 (broad lines) and type 2 (narrow lines) 
AGN, respectively. Based on their hard to soft 
flux ratios, there are approximately equal numbers of type 1 and 
type 2 AGN among the hard X-ray selected 
sources in the LH and EGS with and without mid-IR emission. 
This finding is in contrast
with the local sample of hard X-ray selected AGN of Piccinotti et al. 
(1982) where most of the sources are classified as type 1 AGN
($70-80$\% for AGN at $z<0.12$).

\section{Spectral Energy Distributions}

We have collected optical and near-IR data for all the X-ray/24$\mu$m 
sources in the EGS and LH 
(Crist{\' o}bal-Hornillos et al. 2003;
Wilson 2003). The observations have been band merged 
with the {\it Spitzer} data as described by Le Floc'h et al. (2004)
We used spectroscopic and photometric redshifts for 5 X-ray sources in the LH 
(Lehmann et al. 2001; Mainieri et al. 2002) and 9 sources in the 
EGS (from the Deep Extragalactic Evolutionary Probe   
and Miyaji et al. 2004). For the remaining sources, 
we estimated photometric redshifts where possible from the stellar spectral peak 
at $\lambda_{\rm rest} = 1.6\,\mu$m (see Le Floc'h et al. 2004).

We then classified the X-ray/$24\,\micron$ sources according to the 
shape of their SEDs. Sources that are relatively flat in $\nu f_\nu$ from the optical 
through the mid-IR, resembling the median QSO SED of Elvis et al. (1994), are 
termed type 1 AGN (Fig.~2). Sources that are 
relatively flat in $\nu f_\nu$ at the IRAC and MIPS wavelengths, but 
whose spectra drop toward the blue are consistent with 
being obscured AGN. Although the redshifts 
cannot be estimated well for these two types of objects, the SEDs 
are distinctive and the classification unambiguous.  
Galaxies with decreasing $\nu f_\nu$ in the range $\lambda_{\rm obs} =3.6-8\,\mu$m
and a significant stellar contribution in the optical and near-IR
resemble type 2 AGN or spiral/starburst galaxies, and their SEDs
appear  intermediate between that of Circinus 
 and that of M82 (Fig.~2). The majority of these sources have 
spectroscopic/photometric redshifts
in the range $z=0.2-1.6$.  A few galaxies have increasing
$\nu f_\nu$ for $\lambda_{\rm obs} \ge 3.6\,\mu$m, and look similar 
to local ULIRGs (Fig.~2). Table~1 lists the X-ray properties and 
SED types for those EGS sources 
with well determined SEDs (Fig.~2).

For the 45 X-ray/24$\mu$m sources in our sample with SED type, we find 
that 10 can be classified as pure type 1 AGN SED, 27 as obscured AGN, 
type 2 AGN, or spiral/M82-like SED,  
and 8 as ULIRG-like SED. In the absence of quality X-ray data, the
optical-to-mid-IR SED does not unambiguously identify AGN activity.
This will complicate efforts to identify complete samples
of AGN via optical and {\it Spitzer} photometry.

\begin{figure}
\plotone{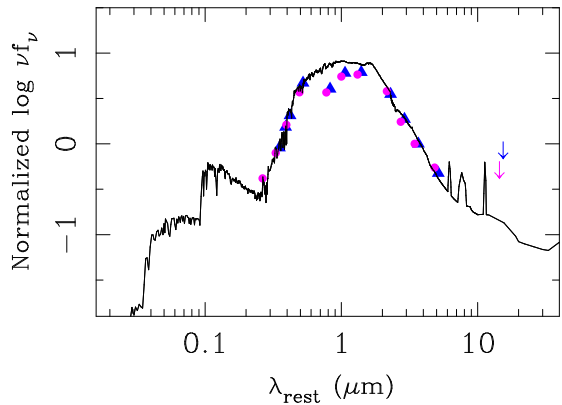}
\caption{SEDs (normalized as in Fig.~2) of two
LH sources --- XMMJ105154.3+572753.2 (circles)
and XMMJ105203.8+572339.3 (triangles)  ---  
not detected at $24\,\mu$m. Their SEDs are  
shown with an S0/Sa galaxy 
template from Devriendt et al. (1999).}
\end{figure}

Fig.~3 shows the hard to soft 
X-ray flux ratio distributions for the three different types of SEDs.  All 
galaxies with type 1 AGN-like
SEDs show a range of hard to soft flux ratios consistent with 
those of spectroscopically classified type 1 AGN
(that is, broad-line AGN). 
Galaxies with obscured AGN, type 2 AGN, and spiral/M82-like
SEDs include a  fraction of sources that would be classified as 
type 1 AGN based on their X-ray properties. ULIRG-SED objects appear to have
a tendency toward softer X-ray flux ratios characteristic of type 1 AGN. 

More than half of the  bright 
X-ray/$24\,\micron$ sources in our sample 
have SEDs dominated by stellar emission or show a significant level of
obscuration. This is consistent with the finding that $40-60$\% of
the Chandra selected galaxies have optical spectra with no signs of nuclear
AGN activity (e.g., Barger et al. 2001; Hornschemeier
et al. 2001).  Many of these galaxies have colors of old stellar populations
(Barger et al. 2003). It is possible that their AGN emission lines
are overwhelmed by stellar light (see Moran, Filippenko, 
\& Chornock 2002).
Others appear highly absorbed in X-rays 
(Barger et al. 2001, see also  Fig.~3); if a similar level
of absorption applies to their emission lines, these lines will
be undetectable. 
In agreement with the finding that 
spectroscopic type 1 (broad lines) AGN behavior 
is relatively rare (e.g., Hornschemeier
et al. 2001), the fraction of type 1 AGN SED dominated 
X-ray sources in our sample is small.

About one-third of the sources with emission in hard X-rays  are not 
detected at $24\,\mu$m. Fig.~4 shows two such X-ray sources in the LH
whose SEDs are consistent 
with being S0 or elliptical galaxies, with no evidence for
the presence of hot dust. Based
on findings by Alexander et al. (2002) for the HDF-N, a significant fraction 
of X-ray sources detected by IRAC but not
by MIPS at $24\,\mu$m are likely to have SEDs like elliptical and S0 galaxies.

To put these trends on a more quantitative basis, we have used local
hard-X-ray selected AGN from Piccinotti et al. (1982) and
Kuraszkiewicz et al. (2003) to construct a comparison sample consisting of
all galaxies with  $z \le$ 0.12 and 
a mid-IR $25\,\micron$ flux density $> 100\,$mJy. In this local sample, 
19 of 32 galaxies, more than half the sample, would be classified by us as type 1 
AGN based on their SEDs. From the
hard X-ray/$24\,\micron$ detections in the EGS and LH, 
we classify 7 of 29 as having SEDs resembling those 
of type 1 AGN.  Since only 2/3 of sources 
with hard X-ray  emission were detected at $24\,\mu$m, some caution
is needed in interpreting this result. However, since a significant fraction 
of the X-ray sources without $24\,\mu$m detections
are likely to have S0/elliptical type SEDs, it is possible that there is
a trend away from pure type 1 AGN behavior with increasing redshift. This possibility
will be probed by further {\it Spitzer} observations of deep X-ray fields that are
currently under analysis.

This work is based [in part] 
on observations made with the Spitzer Space Telescope, which 
is operated by the Jet Propulsion Laboratory, Caltech 
under NASA contract 1407. Support 
for this work was provided by NASA through Contract no. 
960785 issued by JPL/Caltech.

\begin{table*}
\scriptsize
\setlength{\tabcolsep}{0.05in}
\caption{Chandra positions and fluxes, and 
SED types for X-ray galaxies in the EGS with well determined SED types.}

\begin{tabular}{cccccccccccl}
\hline
RA           &  DEC     & d  & $f$(0.5-8keV)   & error   & d  & $f$(0.5-2keV)   & error   & d  & $f$(2-8keV) & error   & SED type\\
(1)          & (2)      & (3)& (4)     &   (5)   &   (6) & (7) &       (8)&       (9) & (10) &        (11)&       (12) \\
\hline
214.189607   & 52.485142 & 1 &  0.244E-13  & 0.237E-14 & 1 &  0.425E-14  & 0.514E-15  & 0 &   0.232E-13  &--  & ULIRG \\
214.333361   & 52.416680 & 1 &  0.573E-14  & 0.114E-14 & 1 &  0.713E-15  & 0.219E-15  & 1 &   0.471E-14  & 0.146E-14  & ULIRG\\
214.486947   & 52.523481 & 1 &  0.809E-14  & 0.129E-14 & 1 &  0.131E-14  & 0.296E-15  & 1 &   0.522E-14  & 0.162E-14  & ULIRG\\
214.355623   & 52.595469 & 1 &  0.535E-14  & 0.133E-14 & 1 &  0.107E-14  & 0.287E-15  & 0 &   0.770E-14  &--  & type 1  AGN   \\
214.222970   & 52.352189 & 1 &  0.196E-13  & 0.299E-14 & 0 &  0.565E-14  & -- & 0 &   0.225E-13  &--  & type 1  AGN           \\
214.312620   & 52.387032 & 1 &  0.177E-13  & 0.197E-14 & 1 &  0.357E-14  & 0.472E-15  & 1 &   0.791E-14  & 0.192E-14  & type 1  AGN \\
214.205410   & 52.425120 & 1 &  0.130E-12  & 0.593E-14 & 1 &  0.258E-13  & 0.140E-14  & 1 &   0.648E-13  & 0.576E-14  & type 1  AGN \\
214.399811   & 52.508273 & 1 &  0.274E-12  & 0.640E-14 & 1 &  0.506E-13  & 0.149E-14  & 1 &   0.127E-12  & 0.554E-14  & type 1  AGN \\
214.401378   & 52.595777 & 1 &  0.657E-14  & 0.140E-14 & 1 &  0.159E-14  & 0.342E-15  & 0 &   0.543E-14  &--  & type 1  AGN         \\
214.395215   & 52.469626 & 1 &  0.342E-13  & 0.293E-14 & 1 &  0.615E-14  & 0.671E-15  & 1 &   0.170E-13  & 0.286E-14  & type 1  AGN \\
214.424483   & 52.473194 & 1 &  0.807E-13  & 0.356E-14 & 1 &  0.141E-13  & 0.880E-15  & 1 &   0.458E-13  & 0.347E-14  & type 2, obscured AGN, Spiral \\
214.441532   & 52.509038 & 1 &  0.243E-13  & 0.203E-14 & 1 &  0.315E-14  & 0.416E-15  & 1 &   0.179E-13  & 0.230E-14  & type 2, obscured AGN, Spiral \\
214.439370   & 52.497597 & 1 &  0.572E-14  & 0.108E-14 & 1 &  0.233E-15  & 0.128E-15  & 1 &   0.856E-14  & 0.165E-14  & type 2, obscured AGN, Spiral \\
214.348256   & 52.432237 & 1 &  0.175E-14  & 0.665E-15 & 0 &  0.421E-15  &--  & 1 &   0.252E-14  & 0.116E-14  & type 2, obscured AGN, Spiral         \\
214.352513   & 52.506880 & 1 &  0.591E-13  & 0.310E-14 & 1 &  0.111E-13  & 0.736E-15  & 1 &   0.316E-13  & 0.292E-14  & type 2, obscured AGN, Spiral \\
214.347510   & 52.531594 & 1 &  0.245E-13  & 0.217E-14 & 1 &  0.582E-14  & 0.576E-15  & 1 &   0.720E-14  & 0.175E-14  & type 2, obscured AGN, Spiral \\
214.267749   & 52.414954 & 1 &  0.132E-13  & 0.168E-14 & 0 &  0.691E-15  & -- & 1 &   0.198E-13  & 0.263E-14  & type 2, obscured AGN, Spiral         \\
214.253100   & 52.322161 & 1 &  0.463E-13  & 0.346E-14 & 0 &  0.137E-13  & -- & 0 &   0.212E-13  &--  & type 2, obscured AGN, Spiral                 \\
214.213653   & 52.346056 & 1 &  0.457E-13  & 0.397E-14 & 1 &  0.871E-14  & 0.854E-15  & 0 &   0.389E-13  &--  & type 2, obscured AGN, Spiral         \\
214.267911   & 52.361115 & 1 &  0.856E-14  & 0.173E-14 & 0 &  0.142E-14  &--  & 0 &   0.181E-13  &--  & type 2, obscured AGN, Spiral                 \\ 
214.294179   & 52.474719 & 1 &  0.145E-13  & 0.179E-14 & 1 &  0.320E-14  & 0.444E-15  & 1 &   0.522E-14  & 0.172E-14  & type 2, obscured AGN, Spiral \\
214.217017   & 52.450224 & 1 &  0.560E-14  & 0.144E-14 & 1 &  0.971E-15  & 0.290E-15  & 1 &   0.442E-14  & 0.184E-14  & type 2, obscured AGN, Spiral \\
214.439939   & 52.460638 & 0 &  0.246E-14   & -- & 1 &  0.101E-14   & 0.136E-15   & 0 &   0.221E-14   & --   & type 2, obscured AGN, Spiral          \\
214.351131   & 52.541650 & 0 &  0.374E-14   & -- & 1 &  0.380E-14   & 0.186E-15   & 0 &   0.231E-14   & --   & type 2, obscured AGN, Spiral          \\
214.345872   & 52.528797 & 1 &  0.589E-14  & 0.117E-14 & 1 &  0.126E-14  & 0.330E-15  & 1 &   0.237E-14  & 0.126E-14  & type 2, obscured AGN, Spiral \\
214.299767   & 52.336922 & 1 &  0.124E-12  & 0.488E-14 & 1 &  0.243E-13  & 0.114E-14  & 0 &   0.714E-13  &--  & type 2, obscured AGN, Spiral         \\
214.313384   & 52.447367 & 1 &  0.650E-14  & 0.125E-14 & 1 &  0.622E-15  & 0.223E-15  & 1 &   0.679E-14  & 0.169E-14  & type 2, obscured AGN, Spiral \\
214.203452   & 52.433096 & 1 &  0.106E-13  & 0.224E-14 & 1 &  0.187E-14  & 0.458E-15  & 0 &   0.135E-13  &--  & type 2, obscured AGN, Spiral         \\
214.390994   & 52.563506 & 1 &  0.778E-14  & 0.130E-14 & 1 &  0.111E-14  & 0.273E-15  & 1 &   0.667E-14  & 0.160E-14  & type 2, obscured AGN, Spiral \\
\hline
\end{tabular}
\smallskip

[This table will be published in the electronic edition of
the Journal.  The printed edition will contain only a sample.]\\
NOTES --- 
In this table we list those EGS galaxies with well determined SED types and with an estimate 
of the redshift, that is, galaxies plotted in Fig.~2. 
Column (1): Chandra RA (J2000).
Column (2): Chandra DEC (J2000).
Column (3): detection in the full ($0.5-8$keV) band (1=dectection, 0=3$\sigma$ upper limit).
Column (4): flux in the full ($0.5-8$keV) band in erg cm$^{-2}$ s$^{-1}$.
Column (5): error of flux in the full ($0.5-0.8$keV) band in erg cm$^{-2}$ s$^{-1}$.
Column (6): detection in the soft ($0.5-2$keV) band (1=dectection, 0=3$\sigma$ upper limit).
Column (7): flux in the soft ($0.5-2$keV) band in erg cm$^{-2}$ s$^{-1}$.
Column (8): error of flux in the soft ($0.5-2$keV) band in erg cm$^{-2}$ s$^{-1}$.
Column (9): detection in the hard ($2-8$keV) band (1=dectection, 0=3$\sigma$ upper limit).
Column (10): flux in the hard ($2-8$keV) band in erg cm$^{-2}$ s$^{-1}$.
Column (11): error of flux in the hard ($2-8$keV) band in erg cm$^{-2}$ s$^{-1}$.
Column (12): SED type (see Fig.~2, and text). 
\end{table*}


\begin{thebibliography}{30}

\bibitem{b1}
Alexander, D. M. et al. 2001, ApJ, 554, 18
\bibitem{b2}
Alexander, D. M. et al. 2002, ApJ, 568, L85
\bibitem{b3}
Barger, A. J. et al. 2003, AJ, 126, 632 
\bibitem{b4}
Barger, A. J. et al. 2001, AJ, 121, 662 
\bibitem{b5}
Boyle, B. J. et al. 1987, MNRAS, 227, 717
\bibitem{b6}
Crist{\' o}bal-Hornillos et al. 2003, \apj,
595, 71 
\bibitem{b7}
Devriendt, J. E. G. et al. 1999,
A\&A, 350, 381
\bibitem{b8}
Egami, E. et al. 2004, in this volume
\bibitem{b9}
Elvis, M. et al. 1994, ApJS, 95, 1
\bibitem{b10}
Fadda, D. et al. 2002, A\&A, 383, 838
\bibitem{b11}
Fazio, G. G. et al. 2004, this volume
\bibitem{b12}
Franceschini, A. et al. 2002, ApJ, 568, 470
\bibitem{b13}
Gilli, R., Risaliti, G., \& Salvati, M. 1999, A\&A, 347, 424
\bibitem{b14}
Gordon, K. et al. 2004, PASP, submitted
\bibitem{b15}
Hasinger, G. et al. 2001, A\&A, 365, L50
\bibitem{b16}
Hornschemeier, A. E. et al. 2001, ApJ, 554, 742
\bibitem{b17}
Huang, J.-S. et al. 2004, this volume 
\bibitem{b18}
Kuraszkiewicz, J. K. et al. 2003, ApJ, 590, 128 
\bibitem{b19}
Le Floc'h, E. et al. 2004, this volume
\bibitem{b20}
Lehmann, I. et al. 2001, A\&A, 371, 833
\bibitem{b21}
Mainieri, V. et al. 2002, A\&A, 393, 42
\bibitem{b22}
Miyaji, T. et al. 2004, AJ, in press (astro-ph/0402617)
\bibitem{b23}
Moran, E. C. et al. 2002, ApJ, 579, 71
\bibitem{b24}
Papovich, C. et al. 2004, this volume
\bibitem{b25}
Piccinotti, G. et al. 1982, ApJ, 253, 485
\bibitem{b26}
Ranalli, P., Comastri, A., \& Setti, G. 2003, A\&A, 399, 39
\bibitem{b27}
Rieke, G. H. et al. 2004, this volume
\bibitem{b28}
Rigby, J. et al. 2004, this volume
\bibitem{b29}
Szokoly, G. P. et al. 2004, ApJS, astro-ph/0312324
\bibitem{b30}
Wilson, G. 2003, ApJ, 585, 191
\end{thebibliography}
\end{document}